\documentclass[english]{article}
\usepackage[T1]{fontenc}
\usepackage[latin9]{inputenc}
\usepackage{babel}
\usepackage{amsmath}
\usepackage{graphicx}
\usepackage[unicode=true]
 {hyperref}



\begin{document}

\title{On the storage and retrieval of primes and other random numbers using n-dimensional geometry }

\author{K. Eswaran}
\maketitle
\begin{abstract}
We show that if you represent all primes with less than n-digits as
points in n-dimensional space, then they can be stored and retrieved
conveniently using n-dimensional geometry. Also once you have calculated
all the prime numbers less than n digits, it is very easy to find
out if a given number having less than n-digits is or is not a prime.
We do this by separating all the primes which are represented by points
in n-dimension space by planes. It so turns out that the number of
planes q, required to separate all the points represented by primes
less than n-digit, are very few in number. Thus we obtain a very efficient
storage and retrieval system in n-dimensional space. In addition the
storage and retrieval repository has the property that when new primes
are added there is no need to start all over, we can begin where we
last left off and add the new primes in the repository and add new
planes that separate them as and when necessary. Also we can arrange
matters such that the repository can begin to accept larger primes
which has more digits say n' where n' > n.

The algorithm does not make use 
of any property of prime numbers or of integers in general,
except for the fact that any  n-digit integer can be represented as a point
 in n-dimension space. Therefore the method can serve to be a storage and retrieval repository
 of any set of given integers, in practical cases they can represent information.
 Thus the algorithm can be used to devise a very efficient storage and retrieval system for
 large amounts of digital data.
\end{abstract}

\section{Introduction}

Mappings of mathematical entities onto points in n-dimensional space
and their separation by planes has been of much interest to mathematicians
{[}1-2{]}. To illustrate our concept,\footnote{This paper has been sent for publication} let us first consider only two
digit numbers: we will represent all two digit numbers as a point
in 2d space following this scheme, the integers 37 will be represented
as a point in two dimensional space: $(x=7,y=3)$ viz $(7,3)$ similarly
the number $43$ will be represented as $(3,4)$. Notice we reverse
the digits from left to right to right to left, we do this so that
larger numbers will be represented as having coordinates of higher
dimension. Similarly the number 1729 will be represented as a point
in 4-dimension space having coordinates $(9,2,7,1)$. 

In the figure below we have represented all the primes below 100 as
points on a graph:\medskip{}

\includegraphics[scale=0.5]{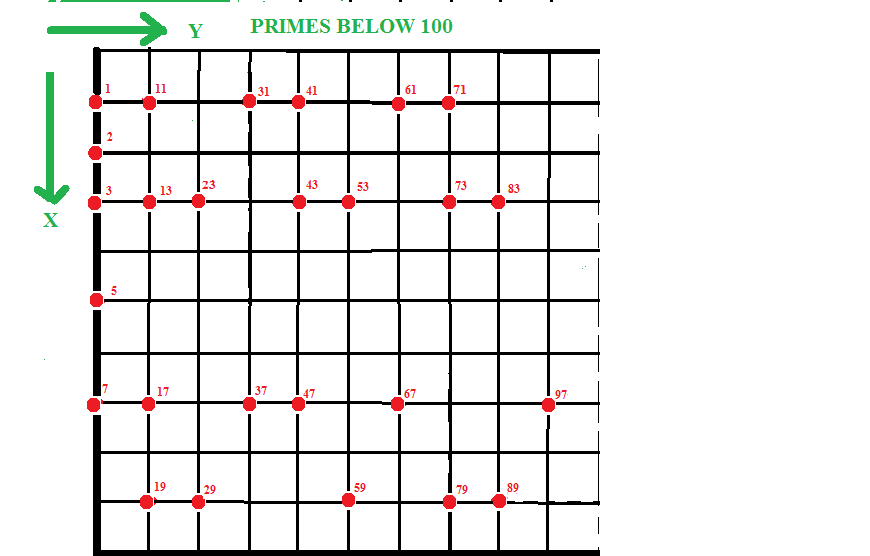}

\medskip{}

Now all these primes can be separated by planes (lines) such that
no two primes will lie on the same side of all planes see figure below:

\medskip{}

\includegraphics[scale=0.5]{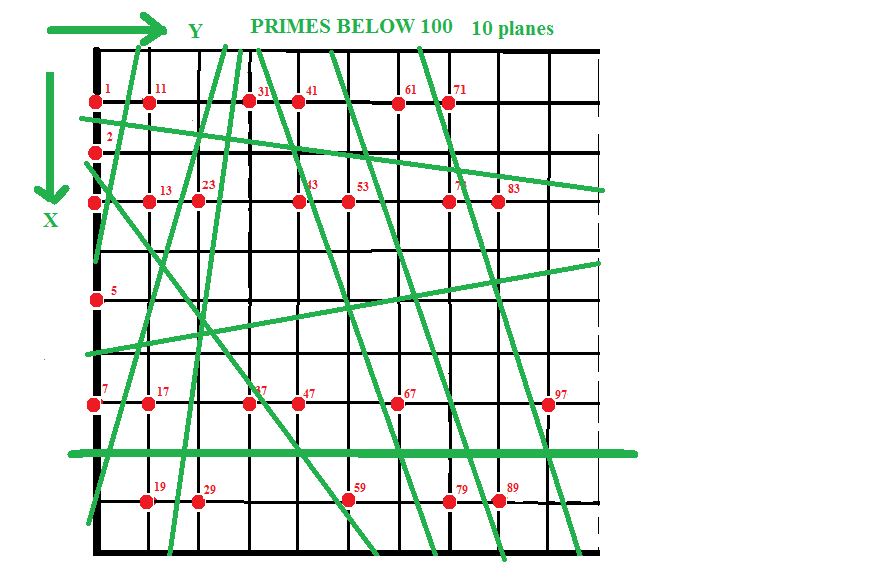}

\medskip{}

We see that only 10 lines are sufficient to separate all the primes.
We can use this information, the graph contains a lot of information,
it indicates that every point representing a prime number is in a

unique position wrt to the 10 planes i.e. no two points are left unseparated
by at least one plane.

\footnote{It may be noted here that the task of finding the planes (lines) of
ALL integers having less than n digits is trivial. For example, in
the case of all numbers less than or equal to 99, we can separate
them by using the 20 coordinate lines $x=1/2;x=3/2;x=5/2;x=7/2;....x=19/2$
and $y=1/2;y=3/2;y=5/2;y=7/2;....y=19/2$%
} This information is stored as a {}``Orientation Vector'' {[}3{]},{[}4{]}
defined for each point (The Appendix contains the definitions of the
terms used in this paper), the dimension of the Orientation Vector
is equal to q, the number of planes present. So if there are two points
P and Q among all the planes that separate the ensemble of points
then by definition $Ov(P)=Ov(Q)$ implies that P and Q are in the
same region wrt all the planes. If $Ov(P)\ne Ov(Q)$ this means P
and Q are separated by at least one plane. The Orientation Vector
of all the points are stored, these are Hamming Vectors we will show
that, using the OV's it is very easy to retrieve any prime number
or determine if any given number is a prime number or not by comparing
the OV's stored.

\section{Brief description of Method}

Suppose we wish to store all prime numbers which have n digits or
below we convert (map) the number to a point in n dimensional space.
Suppose n=5 and we have numbers: $80,917$ and $641$ which are prime
numbers, these are mapped to points which have coordinates$(x_{1},x_{2},x_{3},x_{4},x_{5})$
equal to $(7,1,9,0,8)$ and $(1,4,6,0,0)$ resp. in a 5 dimensional
space. We divide the task into to parts The Storage Task and the Retrieval
Task as follows:

\medskip{}

STORAGE TASK:

1. Create a repository of prime numbers having less than (say) n digits,
represented as points in n-dimension space.

2. Use an algorithm {[}4{]}, which finds the planes that separate
all the points in the repository. (See Appendix, for the definition
of terms: {}``Orientation Vector'' and {}``Position Vector''.)

3. The algorithms finds the number of planes, q, and their coefficients
which define each plane as well as the Orientation Vector of each
Point. We can also store the {}``Position Vector'' (see definition.
The OV are obtained from the Position Vectors.)

\medskip{}

RETRIEVAL TASK

If you are given a n-digit number number R and you do not know whether
it is composite or prime, you do the following: 

1. Find the Orientation Vector of this point R, this task takes $q.n$
operations of multiplications and additions, 

2. Having obtained $Ov(R)$ compare this with all the other OV's stored
in the repository, this involves only bit comparisons. However if
the OV's are sorted in {}``dictionary'' order it does not take many
bit comparisons.) By performing these bit comparisions you will quickly
find that prime number P, whose OV is the same as R i.e. $Ov(P)=Ov(Q)$ 

3. But yet you will not yet know if R is a prime or not, you only
know that it is in the same {}``quadrant'' location as point P.
To ascertain that R is really a prime we need to find its Position
Vector and check if $Pv(R)=Pv(P)$ . If so you will know that R is
nothing but P and is a prime.

(In fact, step 3 is not necessary, if you are storing the prime number
P along with its Orientation Vector in the Repository then all you
need to check if R equals P.This is done if you use the Orientation
Vector of everp prime as an identifier (or ID) and store the prime
number in a repository, next to its Orientation Vector. So if the
OV is known the prime number can be read. And you will soon discover
by consulting the list of OV's, if R is the same as the stored P,
if so, R is prime, if not R is composite.

5. \textbf{Additionally:} Since the algorithm does not make use 
of any property of prime numbers or of integers in general,
except for the fact that any  n-digit integer can be represented as a point
 in n-dimension space. Therefore the method can serve to be a storage and retrieval repository
 of any set of given integers even random integers. 
 In practical cases these integers could represent data or information.
Thus we have an algorithm which serves as a very efficient storage and retrieval system, for digital data, 
which has the  property that at any time new information can simply be added on without restarting, for future retrieval.

\section*{Some Interesting Facts}

1. The method that we use to find the planes separating points in
n- dimension space, does not require you to restart from the beginning
when new points are added and require to be separated both from themselves
and from the existing points. The algorithm permits you to continue
from where you left off.

2. If the dimension of space increases (this will happen when we want
to store primes involving more digits say $m,m>n$), the algorithm
permits you to do this after some initial adjustments you can again
start from where you left off.

3. We make an additional important point: the algorithm does not make use 
of any property of prime numbers or of integers in general,
except for the fact that any  n-digit integer can be represented as a point
 in n-dimension space. All the algorithm needs is that a set G containing $N_f$ points 
in n-dimension space be well defined, and each of their coordinates known before hand.
 The algorithm then proceeds to separate each of the points in G, by planes in n-dimensional space and find the equations of the $q$ planes that do so.
 
4. Therefore the set of points G, can represent any integers, (not just primes), and even
a given set of randomly chosen  integers,  the only specification  is that the set G be well defined. 
The method can be used to separate even real points (whose coordinates may be non-rational).
 
 5. Curiously, if the set G consists only of points whose coordinates are all rational,
 (this will happen if they represent integers according to the scheme above), then matters can be so arranged that
the  $q$ planes which separate all the points, acting like boundaries,  can never pass through 
any point P whose coordinates are all rational, (see footnote 11, just before Appendix B). 
Thus ensuring that points which represent integers will either be on one or the other side of such planes and will never lie on
any one of the $q$ planes, no matter how many points (representing integers) you add on to G later on.

\medskip{}

The above scheme permits one to create an {}``Eternal Repository''
of primes for all time, which can store and retrieve new primes as
and when they are found and at the same time retaining the old.
 \footnote{We can use the method to store and retrieve any set of integers,
eg only twin primes, or or every 100th prime etc. or numbers which are
 solutions to a set of given Diophantine equations etc., the only stipulation is 
that G should be defined in the beginning of the algorithm.}

See the Appendix for the proof and definitions.

\section{Number of Computations required to set up our Prime Number Repository}

We have calculated the number of computations required to set up our
Prime Number Repository for all prime numbers having equal to or less
than $n-$digits is:  

$\approx$ ($n.10^{n+1}$ + $7.O(n^{4}))$ mutiplications, and approx.
an equal number of additions.

And to find if a particular number is a composite or prime, we need
only to make bit comparisons with all the other $N_f$ Hamming Vectors stored
in the data base which works out to be approx: $\sim8.N_{f}$
bit comparisons. Where $N_{f}$is the number of primes stored in
the repository.%
\footnote{We can think of our Prime Repository as some thing like the Sieve
of Eratosthenes, except that we are storing the prime numbers as points
in n-dimension space and retrieving them in a different manner.}

\subsection{Brief summary of Results from a computer program}
  A computer program based on the algorithm was written, it was tried out for 
separating random points in a cube of 15 dimensions.
   Two thousand random points were generated in a unit cube of 15 dimensions, the algorithm
  separated each point by using 22  planes. Details of these and some additional results involving 50,000 points
  in a 25 dimension cube, being separated by 27 planes can be seen in Ref [5].

\section{References}

1. William B. Johnson and Joram Lindenstrauss: Extensions of Lipschitz
mappings on to a Hilbert Space, Contemporary Mathematics, 26, pp 189-206
(1984)

2. Ralph P. Boland and Jorge Urrutia: Separating Collection of points
in Euclidean Spaces, Information Processing Letters, vol 53, no.4,
pp, 177-183 (1995)

3. K.Eswaran:A system and method of classification etc. Patents filed
IPO No.(a) 1256/CHE July 2006 and (b) 2669/CHE June 2015

4. K.Eswaran: A non-Iterative method of separation of points by planes
in n dimensions and its application, Sent for Publication:

 \href{http://www.arxiv.org/abs/1509.08472}{www.arxiv.org/abs/1509.08472}
 
5. K.Eswaran: Results of Separation of Points in high dimension space from a 
Computer Program based on a non-iterative algorithm.

\href{http://www.researchgate.net/publication/285149835}{www.researchgate.net/publication/285149835}
\section{Acknowledgments}

The author thanks the management of Sreenidhi Institute of Science
and Technology for their sustained support. He proclaims his grateful
thanks to his wife Suhasini, and her ever willingness to be a sounding
board on innumerable occasions and to listen to his monologues without
which he does not think this work would have ever been done.

\section{APPENDIX A}

In this Appendix we give the necessary definitions and briefly prove the fundamental
algorithm by which planes which separate a given set of points could
be found. The more detailed proofs are available in K.Eswaran Ref.
{[}4{]} 

Our task is to determine the equations of the planes that can separate
the points in such a way that every point is separated from another
by at least one plane.

Assumption: It is assumed that we have specified a (defined) normal
direction, a point which lies on the positive side of the normal is
said to lie on the positive side of the plane, on the other hand if
the point lies on the other side it is said to lie on the negative
side.This direction is easily found if the equation of the plane is
known for example if 
\begin{equation}
1+\alpha_{1}x_{1}+\alpha_{2}x_{2}+...,+\alpha_{n}x_{n}=0
\end{equation}

is the equation to some $n$- dimensional plane then a point P whose
coordinates are $(p_{1,}p_{2},....,p_{n})$ will be said to be on
the positive side if $1+\alpha_{1}p_{1}+\alpha_{2}p_{2}+...,+\alpha_{n}p_{n}>0$
and in the negative side if

$1+\alpha_{1}p_{1}+\alpha_{2}p_{2}+...,+\alpha_{n}p_{n}<0$

Each component of the Orientation Vector of a point P in X space
Orientation Vector (see definition below is defined by finding out
if P is on the positive side or negative side of each plane.

\subsection{Definitions and Terminology: }

\textbf{Orientation Vector}: Suppose we have a point P in n-dimension
space \textbf{(also called X-space)} and suppose we are given 3-planes.
we define a Orientation Vector as a Hamming vector whose components
precisely specify on which side the point P lies with respect to the
3 planes. Example: if the point P lies on the positive side of plane
1, negative side of plane 2 and positive side of plane 3 , then we
define the Orientation Vector associated with P as $Ov(P)$ and define
$Ov(P)\equiv(1,-1,+1)$. Similarly a point Q which lies on the negative
side of plane 1, negative side of plane 2 and positive side of plane
3 will have an Orientation Vector $Ov(Q)\equiv(-1,-1,1)$ . Points,
whose Orientation Vector differ from another by at least one one component
can be said to be separated by at least one plane.

\textbf{Q Space or Hamming space: }We will also call the space spanned
by the Orientation Vector as Hamming Space or Q space. Since each
point P in X-space has an Orientation (Hamming) Vector associated
with it, we can imagine that all the points in X space are mapped
to a point in Hamming space. Of course this mapping is many to one,
but the point to notice is the following: Let P be some point in X-Space,
then all points, R, in X space which are not separated from P by planes
will all have the same Orientation Vector as P ie. $Ov(P)=Ov(R)$
and we describe this by saying: {}``Points P and R belong to the
same `quadrant''' this statement is certainly true for the {}``images''
of P and R in Q-Space, but we will loosely use this terminology for
the points in X-space and say P and R are in the same {}``quadrant'',
what we really mean is that P and R are points in X-Space and that
they are not separated by planes; Sometimes we will use the term \textbf{neighbors}
and say P and R are neighbors. The point to remember is that P and
R will be neighbors if and only if $Ov(P)=Ov(R)$. Therefore if one
wishes to find out if two points A and B are separated in X-Space
which contains planes, all one needs to do is to compare their Orientation
Vectors: $Ov(A)$ with $Ov(B)$.

\textbf{A Saturated Plane:} We consider a plane in $n$ dimension
space to be {}``saturated'' if it has already been constrained to
pass through $n$ points and hence cannot be adjusted to pass through
a new point, the coefficients of such a plane are completely determinable.
Eg. A plane in 3 dimension gets saturated if it is made to pass through
three points.

We will suppose we are given all the data containing $N_{f}$ points
along with their coordinates in $n$ dimension space are available.

\section*{Steps of the Algorithm}

Step 1: Initially collect a small number of initial points numbering
$N_{0}$ and choose a set of $q_{0}$ planes that separate each one
of these $N_{0}$ points and put these planes in S. The coefficients
defining the equations of these $q_{0}$ planes should be determined
(or randomly chosen but ensuring that they separate the $N_{0}$),
call $N=N_{0}$ and $q=q_{0}$. Store all Orientation Vectors of points
in S in array V. In addition we need a set T which will contain points
which cannot become immediate members of S but are prospective members
and will become members of S eventually. Set T will contain points
which are neighbours of a point which already is a member of S. To
make things simple we will assume that at the start, all the $q_{0}$
planes are saturated. (We do not want to adjust any of these planes)%
\footnote{To avoid `starting troubles', choose $q_{0}$ and $N_{0}$ s.t. $n<2^{q_{0}}$
and $N_{0}>n$%
}. Put Counter = 0.

Step 2: If no more points in G go to step 7, else: Randomly choose
a new point from G, which could be a candidate point to be put in
S, from the remaining points not in S, go to Step 3.

Step 3: Check if this new point is in a new `quadrant', this involves
finding its Orientation Vector and comparing with those in S. If the
point is in a new quadrant, put this point in S and store its Orientation
Vector in V, put $N=N+1$ go to Step 2, if not, it means it has a
neighbor in its quadrant. Go to step 4.

Step 4: (You will come here only if the current point has a neighbor
in S. Notation and procedure for this Step: We keep count of the number
of members of S which have neighbours. Such points are called $a_{i}$,
its first neighbour will be called$b_{i}$, if $a_{i}$ has a second
neighbour this will be called $c_{i}$ and the third will be called
$d_{i}$ )

Put this new point in Set T, $Counter=Counter+1$ define $i=Counter$
call the point with which this new point is a neighbour as $a_{i}$.
If the new point is the first neighbour call it $b_{i}$if $b_{i}$exists
call it $c_{i}$and if $c_{i}$exists call it $d_{i}$ and if $d_{i}$
exists put back the new point in G and go to step 2. Calculate the
mid point of the \textquotedbl{}segment\textquotedbl{}$(a_{i},b_{i})$,
if not already calculated, and call it $m_{i}$, add the coordinates
$m_{i}$, in a {}``List of Midpoints''.

If $Counter=n$ go to Step 5, if less than n, go to Step 2.

Step 5: Since $Counter=n$, this means you have collected n points
in T, each of which have to be separated from their neighbors. However
we, first check if all the $n$ points collected in T are in different
quadrants. There is a slight chance say $b_{i}$ and $b_{k}$ in T
are in the same quadrant, if so, then mark one of them say $b_{k}$
as $c_{i}$, if the latter doesn't exist, else as $d_{i}$ (If both
$c_{i}$and $d_{i}$both exist there is no place for $b_{k}$put it
back in G , restore the count of points in G as well as restore in
either case$Counter=Counter-1$, check similarly for other $b's$
and then go to Step 2.)%
\footnote{We have permitted three points in T to belong to the same quadrant
but not 4 or more. This is just to simplify the algorithm, anyway,
such events are extremely rare. Since, for large $n$, if there are
$q$ planes there are $2^{q}$ quadrants, hence the chances of two
points randomly chosen to be in the same quadrant is $O(1/2^{q})$.
We permit max. no of neighbors,3, just to ensure that even in the
freakiest conditions the algorithm comes to a halt successfully. Of
course the the other possibility is the presence of points of `accumulation',
however this does not happen when points represent integers.%
}

Now in this step we will separate the n pairs collected in T by introducing
a new plane which passes through all the $n$ mid points, whose coordinates
are available in {}``List of Midpoints'' in the array M: {}``List
of Midpoints'' and determine the coefficients of the new plane, by
solving the$n$ constraint eqs. to pass it through the $n$ midpoints;%
\footnote{The coefficients of the plane containing the n midpoints, can be found
by using Gauss elimination, Gram-Schmidt evaluation or by using the
QR algorithm, the last is more useful just in case the n mid points
fall in a plane of n-1 dimension or less, then the rank of the coefficient
matrix will become less than n. This will happen rarely, and even
if it does, it only means we can accomodate another point (or points)
in T with a neighbor in S, thus making the rank n. In such a case
all the points in T (even though they are now more than n) can be
separated by a single plane - the algorithm can proceed.%
} call this plane as plane number $q+1$ and put $q=q+1$, and include
this new plane in S along with all the $n$ points labeled $b_{j},j=1,2..,n)$
and their coordinates, put $N=N+n$,

Step 6: Now check the neighbours of all the $a_{i}$ ,$i=1,2,..,Counter$
if only $c_{i}$ exists then it gets promoted to first neighbour.
But there is a slight complication after $a_{i}$ and $b_{i}are$
separated $c_{i}$can be a neighbour either of $a_{i}$ or $b_{i}$
so if it is the former call $c_i$ as $b_i$ else call $b_i$  as
$a_i$ and $c_i $ as  $b_i$ the same kind of investigation needs
to be done for $d_i $ because it can now belong to the old  $ a_i $ 
quadrant or belong to the new $b_i $ quadrant, in either case the
$d_i$ gets promoted as second neighbour and it will be called $c_{i}$.
Calculate the new mid point $m_{i}$ for the just created segment
$(a_{i},b_{i})$ and store.

After finishing this task, the number, $r$ of second neighbours that
were promoted as first neighbours will be known, after drawing the
plane $q+1$, then call $Counter = r$.

Update V, the Orientation Vectors of points now stored in S wrt to
this new plane (their dimensions become $q+1$); Clear the data in
M: {}``List of Midpoints'' of all points T you have transfered to
S and remove data, of these $k$ points that have been transfered
to S, from set T (most of the time $ k=n$) and go to Step 2.

Step 7: (You will come here only if no more points are left in G $N_{f}=0$
and Counter is not yet $=n$ ) If Counter = 0, Stop else introduce
a new plane passing through the midpoints of segments collected so
far, (T will contain as many points as the value of Counter), since
Counter $<n$, some of the coefficients of this plane can be randomly
chosen; (take action like step 6 to check all these new points are
in their own quadrants), update V, the Orientation Vectors of points
now stored in S wrt to this new plane (their dimensions become $q+1$);
call this plane $q=q+1$, Counter = 0, Clear the data in M: {}``List
of Midpoints'' then if now $N_{f}=0$ i.e. G is empty, Stop; else
go back to Step 2. \textbf{At the end of section 6.2, we mention that
the case when $Counter<n$ and G is empty which is one of the possibilities
in Step 7; the situation can be dealt with in an easier manner.}

END OF ALGORITHM

The algorithm works swiftly most of the time, since in our case all
the point are discrete and there are no limit points or ther is no
{}``accumulation points'' in G the algorithm will always come to
a halt successfully. The general case when the input data is say prepared
by another program , then certain conditions need to be met to ensure
the correct halting of the algorithm. These are idscussed in the cted
reference.

\subsection*{DISCUSSION:}

The following crucial concepts will immediately clarify the steps
of the whole algorithm:: In n dimension space, if we are given n line
segements;

$(a_{1},b_{1}),(a_{2},b_{2}),(a_{3},b_{3}),...,(a_{n},b_{n})$,

Then a single plane passing through the mid points of each segment
will separate the n points $a_{1},a_{2},a_{3},..,a_{n}$ from the
n points $b_{1},b_{2},b{}_{3},..,b{}_{n}$ that is the $a's$ and
the $b's$ will lie on opposite sides of the plane. \textbf{But this
does not ensure that the $a's$ are separate from each other and the
$b's$ are separate from each other.} In order to ensure this we have
imposed the condition that each of the $a's$ belong to Set S and
are \textbf{already separate from each other }and have all have different
Orientation Vectors (in the space of q planes which are in S), \textbf{Similarly
by imposing the condition that each $b$ , (each of which belong to
T), has only one neighbor in S}, (for the moment ignore the second
and third neighbours$c's$ and $d's$) we are making sure that each
$b_{j}$ is separate from other $b's$ though, each $b_{j}$ is not
separate from its neighbor in S namely $a_{j}$. Thus each pair of
points $(a_{j},b_{j})$ have the same Orientation Vector. Now if the
$(q+1$)st plane is drawn then it will ensure that all the $2n$ points
i.e. the set of all the $a's$ and the set of all the $b's$ are not
only separate but are separated from each other. We can then include
plane $q+1$ and the $n$ points viz. the $b's$ in S. This intuitive
result is mathematically easily demonstrable:\textbf{ }

\textbf{Proof:} Since we have included the $(q+1$)st plane, all the
Orientation Vectors have gained one more dimension and have $q+1$
dimensions, the last,$(q+1)$ st component of the Orientation Vector
of point $a_{j}$ will differ from the last component of the Orientation
Vector of its ex-neighbor $b_{j},$ because they are now on either
sides of plane $(q+1$). Thus proving that all the $2n$ points are
now separate\textbf{. QED}

The rare cases of three points belonging to the same quadrant (viz
$b_{i},c_{i},d_{i}$) is permitted but not four, we do this only
to avoid unecessary return of points to G, once randomly chosen from
G. However, one can simplify the algorithm if one just returns the
second point to G and then choose a new point at random.%
\footnote{This simplification of the algorithm will work most of the time work.
But there may be some freakish cases when (say) the last few points
left in G all belong to the same quadrant! This would mean introducing
a few unnecessary number of planes in the end.%
}

\subsection{PROOF OF ALGORITHM:}

At the start S only contains points which are separable from each
other, because that is the way they have been selected. Since $q=q_{0}$
have been selected initially each of the Orientation Vectors of the
points in S, are $q$ dimension vectors. They are all different.

Now till the time we come to step 5 we are just collecting points
which are separable and putting them S or in case thay have a neighbor
in S then we are putting the points in T. This can go on till there
are n points in T and we arrive at Step 5.

At this point there are n points in T and N points in S. Firstly,
there can never be two points in S which are neighbors to a single
point in T. This is because every point in S has a different Orientation
(Hamming) Vectors so they belong to different quadrants in Hamming
space and one point in T cannot belong to two quadrants.

There are now only two possibilities (i) Each point in T has one distinct
neighbor in S or (ii) there are two or three points in T which have
the \textit{same} neighbor in S. We will consider the first possibility
(i): Since by choice, every point in T must have one neighbor in S,
they all belong to different quadrants, since there are n points which
can be separated by the $q+1$ st plane after implementing Step 5.
Then in the new $q+1$ space all the points (the old points in S and
the n new points in T) are all separable form one another and hence
all the n points in T which awere first neighbours the $b_{i}$'s
can now be included in S, we must add the $q+1$ st component to each
Orientation Vector in S to make them into Orientation Vector wrt to
the $q+1$ st plane.

Now coming to the second possibility, if there are two or three points
in T which have the same neighbor in S then after introducing the
new $(q+1)$ plane only one of the points in T namely the one which
is the first neighbour of $a_{i}$say $b_{i}$can be transfered to
S and the other points $c_{i}and$ $d_{i}$need to be kept in T with
one of the points $a_{i}$or $b_{i}$ as its neighbor and the newly
calculated midpoint.(Because the $q+1$ plane has separated the $a_{i}$and
$b_{i}$ they are on opposite sides of plane $q+1$, so we have to
now determine on which sides $c_{i}and$ $d_{i}$ are; the seemingly
complicated arguments are only to ensure that we make the right decision
about $c_{i}and$ $d_{i}$). However, the situation (ii) is rare in
high dimension n space, since we are choosing points in random, and
can best be avoided by not choosing two points in T which are neighbors
to the same point in S (this means ejecting the second point from
T and putting it back among $N_{f}$ in G, or by keeping it in T and
use it, later, only after you have introduced a new plane or even
after a few new planes.) In brief, the situation (ii) can be avoided
though it just causes difficulties in programming.

So we see after step 5 and Step 6 we will have an enhanced set S with
more points and one more plane and all of which are separated by these
$q+1$ planes. So we can call $(q+1)$ as $q$ and then re-do steps
2 to 6 till all points $N_{f}$ in G are exhausted. Then S will contain
all the points and all the planes which separate them.

Step 7 will happen in the end when all the points are exhausted but
there are less than\textit{ n }points in T and these must be separated.\textbf{
}The logic of Step 7 is similar to step 6\textbf{.}

\textbf{It may be mentioned here that Step 7 can be completely avoided.}
If we find that there are say less than n points in T. Let us say
r is a number such $Counter+r=n$. All we have to do is choose r random
points (these should not have been in G) each of which has as a neighbor
in its quadrant, one of the points which was drawn from G but now
in S. Then put each of these r points along with its neighbor as a
pair in T. Since you have already $Counter$ number of pairs in T,
with the addition of these r pairs, we have $n$ pairs. Now since
$Counter=n$, we can choose the last plane separating all the $n$
pairs.%
\footnote{These are just `End-Game', tricks that can make programming simpler%
}

So we see at the end of the algorithm S will contain all the points
along with the equations of the $q$ planes that separate them,\textbf{
}which are the needed outputs. \textbf{QED} %
\footnote{It may be noted that this algorithm is non iterative, and every point
is mostly {}``looked at'' only once.%
}

In the above algorithm it is assumed that every new plane does not
`split' some other point (pass through). However if this happens (it
is a very rare event), to some point, we will have to shift the `mid
points' of the segments slightly to one side (because to cut a segment
into two parts it is not necessary that the plane passes exactly though
its mid point) so that the plane passes to one side of the offending
point, %
\footnote{Instead of shifting a point already in S, we could shift all the mid points
by a small distance $\delta$ in the direction of the new normal
and recalculate the plane,thus moving it to one side
of the point upon which it was formerly incident. We do this so the algorithm won't fail. 

Another neat trick
which, pure mathematicians who are disciples of Cantor, may admire, is to replace the constant 1 by $\tau$
 (this 1 occurs in every  equation
 which defines the q  planes in Eqs., (1), (2),..., (q) below); and choose $\tau$ to be a transcendental number which is almost unity say choose $\tau = \pi/ \pi_{20}$, where $\pi_{20}$ is the value of $\pi$ correct  to 20 decimal places.
Then we can be assured that the plane can never pass through a point which is represented by integer coordinates,
 this is especially true since all the coefficients $\alpha_{i,j}$ are rational, being obtained by solving a finite set of equations whose coefficients are rational (because all the mid points are means of two rational numbers). This replacement of 1 by $\tau$ (which is very close to unity) should be done only after we have obtained all the $\alpha_{i,j}$,  we thus doubly ensure that all points which represent prime numbers will never lie exactly on any of the q planes. Strange as this may seem we have, succeeded in separating all prime numbers in n-dimensional space by using q planes, which act as boundaries which contain no rational but only algebraic and transcendental points!} and then proceed with the algorithm.

\medskip{}

\section{APPENDIX B}

\subsection*{(A) Finding Orientation Vector of a point P among q planes}

\textbf{To find the Orientation Vector }of a particular point P whose
cords are $x_{1}=p_{1},$ $x_{2}=p_{2},...,x_{n}=p_{n})$ that is$(p_{1},p_{2},..,p_{n})$
in each of the q linear equations that define the q planes.

Suppose the Equation to the first plane is:

\begin{equation}
1+\alpha_{11}x_{1}+\alpha_{12}x_{2}+....+\alpha_{1n}x_{n}=0
\end{equation}

the second equation 

\begin{equation}
1+\alpha_{21}x_{1}+\alpha_{22}x_{2}+....+\alpha_{2n}x_{n}=0
\end{equation}

\[
---------
\]

and so on til the last qth equation 

\[
1+\alpha_{q1}x_{1}+\alpha_{q2}x_{2}+....+\alpha_{qn}x_{n}=0\quad(q)
\]

And you substitute the coord of point P in the above equations, obviously
you will not get zeros on the rhs because we assume P is not on any
of the planes and therefore they do not satisfy the equations. So
if you Substitute you will get . 

\begin{equation}
1+\alpha_{11}p_{1}+\alpha_{12}p_{2}+....+\alpha_{1n}p_{n}=r_{1}
\end{equation}

the second equation 

\begin{equation}
1+\alpha_{21}p_{1}+\alpha_{22}p_{2}+....+\alpha_{2n}p_{n}=r_{2}
\end{equation}

\[
---------
\]

and so on till the last qth equation 

\[
1+\alpha_{q1}p_{1}+\alpha_{q2}p_{2}+....+\alpha_{qn}p_{n}=r_{q}\quad(q)
\]

We define the Position Vector of point P wrt to q planes as :

\begin{equation}
Pv(P)=(r_{1},r_{2},r_{3},.....r_{q})
\end{equation}

Now we define the Orientation Vector of Point P as a Hamming vector
which has exactly q components.

\begin{equation}
Ov(P)=(h_{1},h_{2},h_{3},.....h_{q})
\end{equation}

Where the $h's$ are numbers which are either +1 or -1. this is defined
by using the Sign function $Sgn(z)=+1,\; if\; z>0$and $Sgn(z)=-1\; if\; z<0$
Sgn(z) . We define them as follows $h_{1}=Sgn(r_{1}),\: h_{2}=Sgn(r_{2}),\: h_{3}=Sgn(r_{3}),\:...\:,h_{q}=Sgn(r_{q})$
.

So we see the Orientation Vector of Point P is a Hamming Vector.

\subsection*{(B) To Calculate the Computational Complexity of the Algorithm:}

\textbf{The calculation of the Hamming Vector $Ov(P)$ for a single
point point P involves precisely $nq$ multiplications, $nq$ additions
and q determinations of sign.} If there are $N_{f}$ points in the
input sample G and finally after the running of the algorithm if we
find that there are eventaually $q_{f}$ planes needed to separate
all the $N_{f}$ points. Then the necessary calculations for determining
all the Orienatation Vectors of all the $N_{f}$ points.is precisly:
\textbf{$N_{f}.nq_{f}$ multiplications, $N_{f}.nq_{f}$ additions
and $N_{f}q_{f}$ determinations of sign. }

The citations {[}Non-Iter{]} refers to Ref {[}4{]}.

\textbf{The calculations necessary to obtain the equations of all
the $q_{f}$ planes is done in 2.6 }{[}Non-Iter{]},\textbf{ as well
as the number of comparisions necessary to detemine if a new point
(which is randomly picked from G) belongs to a new quadrant in S or
to an old quadrant requires Hamming vector comaprisions with all the
OV's of points already in S. All this is done in 2.6, opcit.}

\medskip{}

The computations involved in this algorithm can be worked out in a
fairly straight forward manner:

1) First of all the planes are not known at the start of the lagorithm. 

2) The equation to each plane (say the jth plane) will be determined
at some stage in the algorithm only when it is requird in S for example
to go through the n mid points of n pairs at some stage in the algorithm
. This process needs a the solutionof $n$ linear equations to determine
the $\alpha_{ji},i=1,2,....,n$ coefficients of the $j^{\{th\}}$
plane.:

\[
1+\alpha_{j1}x_{1}+\alpha_{j2}x_{2}+....+\alpha_{jn}x_{n}=0
\]

This computation takes involves a Gauss elimination process and involves
$(2/3)O(n)^{3}$ multiplications.Since there are eventaully
$q_{f}$planesthe multiplications involved are $(2/3).q_{f}.O(n)^{3}$
and if by assumption $q_{f}=O(log_{2}(N))$ ,(this is only true when for large dimension space ($2^n > N$)) we can get an estimate when n is large.

4. The Hamming Vector is Unique. if there are two points P and Q in
S then $Ov(P)=Ov(Q)$ implies Pand Q are in the same quadrant if not
they are in different quadrants. This property is used time and again
in the algorithimic process to determine whether every new entrant
qualifies to be in S or in T (the set of temporary members).

3. Every time a new point enters S its Hamming Vector has to be computed
and then compared with Hamming Vectors of all the existing points
already in S, suppose there are at this stage N points in S and q
planes. The Hamming vector will be of dimension q thus there will
be $N.q$ bit comparisions. Of course ALL these comparisions are not
necessary as stated in the paper (We can quit the comparision early
when we find a bit difference).Eventually at one time or other, in
the algoritmic process, the Hamming vector of each point is compared
with the Hamming vector of all the other $N_{f}-1$ points leading
to $N{}_{f}^{2}$ comparisions.

5. Note every time a new plane is introduced in S the dimension of
the Hamming Vector of all the points in S increases by one, so it
has to be updated. so the rhs of the equation of the new plane need
to be capculated for each point in S. That is $r_{q+1}$ needs to
be calculated for each N in S, thus there are $N.n$ multiplications
to determine the $q+1$st bit for each of the N Hamming Vectors.

6. All the above points have been taken account in Sec 2.6{[}Non-Iter{]}

7. After the algorithm runs its course all the coordinates of the
points in Sare unchanged and they are exactly the same as they had
in G. S and G are mirror spaces. Except that finally S contains $q_{f}$
planes in addition to the $N_{f}$ points. All the coefficients of
the equations defining each of the $q_{f}$ planes are precisely calculated.

This is how we arrived at the computational complexity of the algorithm
which is

1. $(2/3).q_{f}.O(n^{3})$ multiplications for generating the
equations of all the planes,

2. $N_{f}q_{f}.n$ multiplications for finding the OrientationVectors
of all the $N_{f}$ points.

3. $\sim8.N_{f}^{2}$ bit comparisions

As was mentioned above for very large n-dimension space when
$2^{n}>N_{f}$ then a reasonable estimate, {[}2{]},{[}4{]}, for the
number of for the number of planes $q_{f}$ is $q_{f}=O(log_{2}(N_{f})),$.

But for our problem of primes which are having $n$ or less than $n$
digits, we need to estimate $q_{f}$ properly. Now we need to estimate
the number of planes required for separting an $n-$digit integer
by planes. We had already seen that all $2-$digit integers can be
separated by 20 planes (these ar nothing but 10 parallel lines parallel
to the x and y axis). So when we have $n-$digit integers we can definitely
separate it by using $10n$ planes (ie we would have 10 planes for
each coordinate), this separates all the n-digit integers but the
number of primes is far less than this. So we can consider $10n$
as the upper bound for $q_{f}$. And the estimate for the number of
primes of n-digits is $N_{f}\approx10^{n}/n.$ (we use the formula
from the prime number theorem which estimates the number of primes
below $N$ as $N/log(N)$.) 

So we see the number of computations required for setting up our Repository
of Prime numbers having n-digits and lower is:

$\approx7n.O(n^{3})$ + $n.10^{n+1}$ multiplications i.e 

$\approx$ ($n.10^{n+1}$ + $7.O(n^{4}))$ multiplications.

------

\medskip{}

\end{document}